# Comment on hep-ph/0603166, hep-ph/0606291, and hep-ph/0606310 by S.S. Afonin

Recently, S. Afonin has published three interesting papers devoted to chiral symmetry breaking and restoration, parity doublets, multispin-parity clusters, and Regge trajectories (RT), [1-3]. This research was mostly targeted towards light unflavored mesons, but baryon resonances were also touched upon slightly. One of Afonin's major points was to relate light mesonic spectra to linear RT's in the *whole* energy interval, i.e.:

$$m^2(n) = an + m^2 \qquad \text{eq. (6) from [2]}$$

$$m^2(n) = a(n + \tfrac{1}{2}) \qquad \text{eq. (21) from [2]}$$

$$m^2(n) = a[n + x + \varepsilon\delta(n)] \qquad \text{eq. (37) from [2]}$$

$$m_{LS}^2(n, J) = 2m_\rho^2(n + J - 1/2) \qquad \text{eq. (46) from [2]}$$

$$m_{AVW}^2(n, J) = 2m_\rho^2(n + J) \qquad \text{eq. (47) from [2]}$$

Of course, everyone remembers formulae like these from their University studies. But our published results [4, 5] clearly overrules forty plus years of dogma of the "straight and parallel RT's" started from Regge's papers [6, 7]. The results of our string and potential model fits and predictions for baryon and meson spectra and RT's reveal a prominent feature – RT in many cases are nonlinear functions of $J$, $L$, and $n_r$. This fundamental feature is in accord with analysis of pure experimental RT's, and with predictions of different quark models, reviewed in [4, 5]. RT's for mesons and baryons are not straight and parallel lines in general in the current *resonance* region, both experimentally and theoretically, but very often have appreciable curvature, which is *flavor*-dependent.

It is well known that string-like models naturally describe large $L$ and $J$ behavior of the hadronic spectra. Why Afonin thinks that *whole* hadronic spectrum should be *linear* and described by the string models, remains to be a puzzle. In our paper [4], it has been shown that realistic massive string model lead to some nonlinearity, and only at $J \gg 1$ the linear spectrum would be established.

Another close by issue in [1-3] is the usage and principal meaning of the WKB-approximation. Author claims in [3] that Eq. (5) should be treated in quasiclassical approximation.

$$M^2(I, G, P, C, L, J, n) \cong a(n + J) + b$$

But the author forgets that the WKB-method always means large $n_r \gg 1$, and *finite J, L*. Actually, the WKB-type model ("h-expansion") was successfully used in the 1990's for mesonic RT, and it clearly leads to nonlinear RT and spectra in the current resonance energy region [8]. Authors [8] proved that nonlinearity would hold for any power-law potential. Also, they show that approximate linearity of the RT in the resonance region really has *nothing* in common with asymptotical linearity, which begins at the mass value $M \approx 7\text{-}8$ GeV, unachievable for light mesons [8]. In paper [9], Brau in yet another WKB-type model came up with a clearly nonlinear regime for mesonic spectra and RT via analytic formulae.

Now, we come to the most important part of this Comment. We are talking about the *clustering feature* of the hadronic spectra. The author claimed that meson clusters are very similar to baryon ones [1], but didn't prove this at all. In what sense they are similar, if baryons have a different classification scheme, and there are ten times more meson species, even in light unflavored sector? Afonin wrote, that clusters occur at 1350 ± 120, 1720 ± 90, 2000 ± 70, 2300 ± 60 MeV [1]. Here we have big mass uncertainties, and the author didn't clarify how they were evaluated. For example, if one will estimate just with low- and high-edge cluster masses, it is easy to get 0.38 – 1.58 GeV$^2$ mass gaps between four clusters, and there is 1.58/0.38 = 4.16 fluctuation ratio with huge dispersion, σ. The author [1] has changed cluster masses in an updated version, compared to original version without an explanation. Further, we see on page 6 [1]: "However these clusters are only multispin ones. What actually happens in reality (at least in considered channels) is multispin-parity clustering $M^2(P,n,J) = M^2(n+J)$." I have to completely disagree with this statement: spin-parity clusters were defined via Lorentz group O(1,3)$_{LS}$ and Rarita-Schwinger multispinors in baryons [10]. N, Δ, Λ, Σ - resonances form "spin" and "parity" clusters, which are not equidistant in mass, or (mass)$^2$. Author's clusters in Fig.1 [2,3] are *completely* different, and grouped as (mass)$^2$ only. I think the classification into clusters should not be performed solely on the grounds of the mass degeneracies. This is because extracting the masses as peak positions of homemade Breit-Wigner parameterizations is quite unreliable. As Hoehler said in a private conversation, the numbers in the brackets are nothing but "names" for the resonances and should not be confused with masses. In order to introduce a cluster idea, one has to have some theoretical guiding scheme. According to Kirchbach [10], the guiding scheme are the representations of SU(2)$_I$ x O(1,3)$_{LS}$ group. Cohen-Glozman had suggested considering chiral multiplets, based on SU(2)$_L$ x SU(2)$_R$ representations [11], and Klempt [12] used SU(6)$_{SF}$ x O(3)$_L$ group. The author [1-3] didn't use any classification scheme. I have serious concerns about cluster masses averaging in [1-3], as it should be weighted with spins by formula like this;

$$M_{COG} = \Sigma_i M_i (2J_i + 1)/( \Sigma_i (2J_i + 1)).$$

In my research I came up with somewhat different clusters structure. I think there is no sense to put "all eggs in one basket" today. We should start from U(1)$_A$, SU(2) left and right groups, going to higher and wider symmetry groups, and construct multiplets and clusters correspondingly. In particular, my findings will give vector-axial vector parity doublets ($\rho_1$ - $a_1$), ($\rho_2$ – $a_2$), ($\rho_3$ – $a_3$), ($\rho_4$ – $a_4$), which can be denoted as a cluster $\rho$ - $a$ (2186 – 2310) MeV. We can combine this with U(1)$_A$ pairs: $\eta$(2190) – $f_0$(2197), $\eta$(2303) - $f_0$(2329), $\eta_2$(2258) – $f_2$(2231), $\pi_4$(2250) – $a_4$(2280), which is cluster (2190 – 2329) MeV. As a result we have a high-lying *supermultiplet* exhibiting a perfect match between vector-axial vector parity doublets and U(1)$_A$ pairs. This clearly shows the road to the total degeneracy and restoration of the chiral symmetry high in the spectrum. I will report all my results soon.

| ρ - a | ⊕ | η π a f | ⇒ | ρ η π / a f | (2186 - 2329) |

Averaged mass for such a cluster would differ from [1-3] a bit.

Author has used different sets of light mesons at Fig.1 in [1] versus [2,3] without explanation. The following states were missed in [1], and present in [2,3]: ω, $h_1$, $b_1$, $\omega_2$, $\omega_3$, $\omega_5$, $h_3$, $b_3$, η, $\eta_2$, $\eta_4$.

As we see from Bugg's review [13] on light mesons, he has quite different clusters: 1590 -1700 MeV, 1930 - 2100 MeV, and 2240 - 2340 MeV, with averages of 1645, 2015, and 2290

MeV, as opposed to 1350, 1720, 2000, and 2300 MeV by Afonin. Author of [1-3] claims the existence of an extra low-lying cluster at 1350 MeV, without serious explanation for it.

In [3] the author didn't clarify, when the conformal symmetry will be restored. The scales of restoration of chiral, $U(1)_A$ and conformal symmetries could be different. In my opinion, the author [1-3] has made numerous interesting observations, but *missed* hitting the physics.

A.E. Inopin
Department of Experimental Nuclear Physics
Kharkov National University
Svobody Sq.4, 61077, Kharkov, Ukraine.